# Probing phonon dynamics with multi-dimensional high harmonic carrier envelope phase spectroscopy

*Ofer Neufeld[1,\*,†], Jin Zhang[1,†], Umberto De Giovannini[1,2,3], Hannes Hübener[1], and Angel Rubio[1,3,4,\*]*

[1]*Max Planck Institute for the Structure and Dynamics of Matter and Center for Free-Electron Laser Science, Hamburg, Germany, 22761.*

[2]*IKERBASQUE, Basque Foundation for Science, E-48011, Bilbao, Spain.*

[3]*Nano-Bio Spectroscopy Group, Universidad del País Vasco UPV/EHU, 20018 San Sebastián, Spain.*

[4]*Center for Computational Quantum Physics (CCQ), The Flatiron Institute, New York 10010, NY, USA.*

[†]*These authors contributed equally.*

\*Corresponding author emails: ofer.neufeld@mpsd.mpg.de, angel.rubio@mpsd.mpg.de.

**Abstract:** We explore pump-probe high harmonic generation (HHG) from monolayer hexagonal-Boron-Nitride, where a terahertz pump excites coherent optical phonons that are subsequently probed by an intense infrared pulse that drives HHG. We find, through state-of-the-art *ab-initio* calculations, that the structure of the emission spectrum is attenuated by the presence of coherent phonons, and is no longer comprised of discrete harmonic orders, but rather of a continuous emission in the plateau region. The HHG yield strongly oscillates as a function of the pump-probe delay, corresponding to ultrafast changes in the lattice such as bond compression or stretching. We further show that in the regime where the excited phonon period and the pulse duration are of the same order of magnitude, the HHG process becomes sensitive to the carrier-envelope-phase (CEP) of the driving field, even though the pulse duration is so long that no such sensitivity is observed in the absence of coherent phonons. The degree of CEP sensitivity *vs.* pump-probe delay is shown to be a highly selective measure for instantaneous structural changes in the lattice, providing a new approach for ultrafast multi-dimensional HHG-spectroscopy. Remarkably, the obtained temporal resolution for phonon dynamics is ~1 femtosecond, which is much shorter than the probe pulse duration because of the inherent sub-cycle contrast mechanism. Our work paves the way towards novel routes of probing phonons and ultrafast material structural changes and provides a mechanism for controlling the high harmonic response.

## Introduction

High harmonic generation (HHG) has recently been established in condensed matter as a source of coherent extreme ultraviolet radiation(1–3), as well as a useful probe for various material properties such as band structure(4, 5), symmetry(6–8), berry-phases(9), topology(10–15), and more. The extremely nonlinear nature of the process can act as a unique platform with enhanced sensitivity for ultrafast dynamics on femtosecond and sub-femtosecond timescales, which has primarily been applied for exploring driven electron dynamics(16–20). In contrast, HHG-spectroscopy as a probe for phonon and lattice dynamics has not been thoroughly studied. It was only recently shown for lower-order 3$^{rd}$ and 5$^{th}$ harmonics that HHG can be sensitive to coherent phonons dynamics(21), though the different physical coupling mechanisms have not yet been identified or analyzed. Here HHG could ideally be employed as an ultra-sensitive probe for ultrafast phonon motion with a plethora of possible physical applications, including lattice-induced symmetry breaking(22), phase transitions(21, 23–26), and energy transfer through non-radiative processes(27).

One common notion is that phononic effects should be quite small in solid HHG, because these would average-out over many lattice unit-cells, and because phonons typically propagate on picosecond timescales that are much longer than the femtosecond dynamics that standardly drive the HHG process. This notion is also supported by HHG experiments in the gas phase, where molecular vibrations have been shown to only weakly affect HHG(28–30) (causing yield modulations smaller than ~10%). Nonetheless, driving coherent phonons in a pump-probe set-up does not necessarily follow the above rule of thumb, since: (i) if the driven dynamics are coherent, their effects would not necessarily average out, and (ii), solid-phases offer a much larger wealth of phononic band excitations than molecules, where the different underlying HHG mechanisms could lead to enhanced nonlinear responses.

Here we theoretically investigate optical phonon-assisted HHG in a pump-probe set-up. Monolayer hexagonal-Boron-Nitride (hBN) is coherently pumped with THz light that initiates strong lattice vibrational motion, which is thereafter probed by intense infrared (IR) pulses that generate high harmonics. The electronic and ionic dynamics are described by state-of-the-art *ab-initio* methods, and analyzed with respect to the pump-probe delay. In the presence of coherent phonons, the HHG spectrum is comprised of a continuous emission rather than a well-defined frequency comb. This effect results from the introduction of the long-duration phonon timescale to the system that breaks the time-translational symmetry of the IR pulse, and vanishes if phonons are assumed to have random phases (as occurs in the thermal case). We show that the HHG yield strongly, and periodically, oscillates with the pump-probe delay, which directly correlates to the instantaneous structural changes in the lattice such as bond compression or stretching (that are inherently connected to the instantaneous electronic structure). Moreover, the phonon motion results in strong carrier-envelope-phase (CEP)-sensitivity in the HHG yield that is absent in the non-pumped system, and which standardly does not appear from such multi-cycle driving(31–35). We demonstrate that the degree of CEP sensitivity *vs.* the pump-probe delay is a very selective measure for femtosecond structural dynamics, which opens new routes for HHG-spectroscopy with sub-cycle temporal resolution.

**Results and Discussion**

We begin by describing the simulated set-up and theoretical approach. Monolayer hBN is described using density functional theory (DFT) within the Kohn-Sham formulation and the local-density approximation (LDA), with two in-plane periodic dimensions, and an out of plane non-periodic axis. All calculations are performed within the open-access real-space based octopus code(36–38). Technical details can be found in the materials and methods section, while specific details about the laser pulse and phonons involved are outlined below. hBN is assumed to interact with a pump THz pulse that excites coherent phonon dynamics, which are subsequently probed by an intense IR laser pulse that generates high harmonics (see Fig. 1 for illustration). The phononic excitation is approximated here by directly initiating the lattice motion rather than describing the interaction with the pump THz pulse, which substantially reduces the computational load. The ionic motion is described by Ehrenfest dynamics(39), where forces are calculated according to the interactions of ions with the laser, the surrounding electron density, and nearby nuclei. The electronic degrees of freedom are described fully quantum mechanically within time-dependent DFT (TDDFT) in the adiabatic approximation(40), in the velocity gauge, and in the dipole approximation. Pseudopotentials are employed to deal with deep core states(41) (see the materials and methods section for details). Notably, this state-of-the-art technique fully incorporates electron-phonon and phonon-phonon couplings, and is thus suitable for studying this complex system.

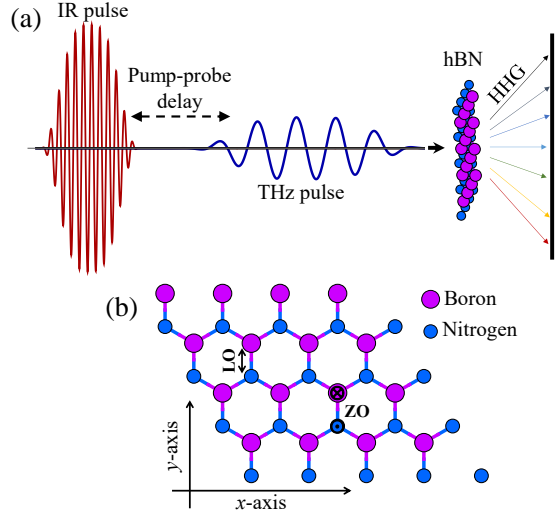

**Figure 1.** Schematic illustration of the pump-probe HHG set-up. (a) A THz pulse excites coherent phonon dynamics in hBN, which are subsequently probed by an intense IR pulse that is polarized in the hBN plane and drives HHG. The harmonic yield is measured with respect to the pump-probe delay, and optionally also with respect to the IR pulse CEP and probe laser polarization. (b) Illustration of hBN structure and explored phononic landscape. Ionic vibrations along the *y*-axis represent LO modes, while motion out of plane represents ZO modes. The HHG yield is modulated due to changes in the lattice structure. For the LO modes, when the B-N bonds along the *y*-axis are compressed, the bonds at 120 degrees from them (almost aligned with the *x*-axis)) are stretched, and vice-versa.

In this pump-probe set-up, there are two main degrees of freedom of interest: the properties of the excited phonon modes and the properties of the HHG driving probe laser pulse. Depending on these, the dynamics and observables can greatly vary. We study here the excitation of in-plane longitudinal optical (LO) phonon modes, and out-of-plane optical (ZO) modes, which have slightly different physical properties (see illustration in Fig. 1(b)). The LO mode has a period of 25.7 femtoseconds (fs) ($\Omega_{LO}$=38.9 THz), while the ZO mode has a slightly longer period of 41.1 fs ($\Omega_{ZO}$=24.3 THz). Both of these modes are excited with amplitudes in the range of a few percent with respect to the lattice parameter and B-N bond length, which is experimentally feasible(42) and still leads to large and measurable effects. We also note that the phonon landscape is naturally anharmonic in hBN(43), which will slightly manifest in the results. The IR probe laser that drives HHG is assumed to have the following vector potential:

$$\mathbf{A}(t) = A_0 f(t)\cos(\omega t + \phi_{CEP})\hat{\mathbf{e}} \quad (1)$$

where $A_0$ is the amplitude of the vector potential that is connected to the amplitude of the applied electric field, taken here as 0.27 V/Å (equivalent to $10^{12}$ W/cm$^2$). This laser power is sufficient to induce extremely nonlinear HHG responses from hBN above the band gap, but remains below the material damage threshold as is required for time-resolved spectroscopy(44, 45). $\hat{\mathbf{e}}$ in Eq. (1) is a linearly-polarized unit vector, *f(t)* is a dimensionless envelope function with a duration of ~25 fs (see the materials and methods section for details), and ω is the fundamental carrier frequency that corresponding to a wavelength of 1600nm. The temporal characteristics of $\mathbf{A}(t)$ are crucial, because they outline a physical regime where the pulse duration is on the same order of magnitude as the phonon period. Moreover, a single period of the carrier wave has a duration of ~5.3 fs, which is on the same order of magnitude as a quarter-cycle of a phonon motion. Altogether, these conditions should lead to a strong modulation of the HHG response, because within the IR pulse duration the lattice undergoes less than a single cycle of ionic displacements, which breaks the time-translation symmetry that is exhibited by the phonon-free laser-matter system. On the other hand, the phonon motion is not considerably slower than changes in the laser envelope, which means that the dynamics cannot be described by a static approximation for the lattice. It is also noteworthy that for these parameters the IR pulse is a multi-cycle pulse that usually does not lead to CEP sensitivity in the HHG spectra ($\phi_{CEP}$ in eq. (1) denotes the CEP that is taken as zero unless stated otherwise). We will show below that contrary to the standard case, such sensitivity is induced due to the presence of coherent phonon modes.

Having outlined the set-up, we explore HHG with a pumped LO phonon mode. Figure 2(a) shows exemplary HHG spectra from the phonon-pumped (red) and phonon-free (blue) systems, respectively. Before addressing the dynamical response, it is worth analyzing the general structure of the spectra: in the absence of phonon dynamics, sharp harmonic peaks are obtained for both even and odd harmonics (because hBN is not inversion-symmetric). The phonon-pumped system on the other hand shows a quasi-continuous plateau from 4 to 12 eV (note that small peaks in the spectra in this region do not appear at integer harmonic orders of the IR probe pulse), while perturbative emission up to 4$^{th}$ order remains well-defined. We also note a sharp peak resonant with the phonon frequency (at ~0.15 eV) which is attributed to the phonon motion and the Born-Oppenheimer dynamics. The lack of frequency-comb harmonics in the higher-order response indicates that the temporal dynamics is not periodic due to the phonon-induced lattice displacement, as expected. The perturbative response on the other hand is more mildly modulated. Interestingly, if the HHG response is integrated over the pump-probe delay, the spectrum regains its frequency-comb nature with well-defined harmonic orders that are very similar to the phonon-free response (see Fig. 2(b)). This result is equivalent to the thermal limit, where different regions in the lattice have random phonon phases that average out. It validates that in standard non-coherent conditions, it would be very difficult to determine experimentally the contributions of phonons to the response.

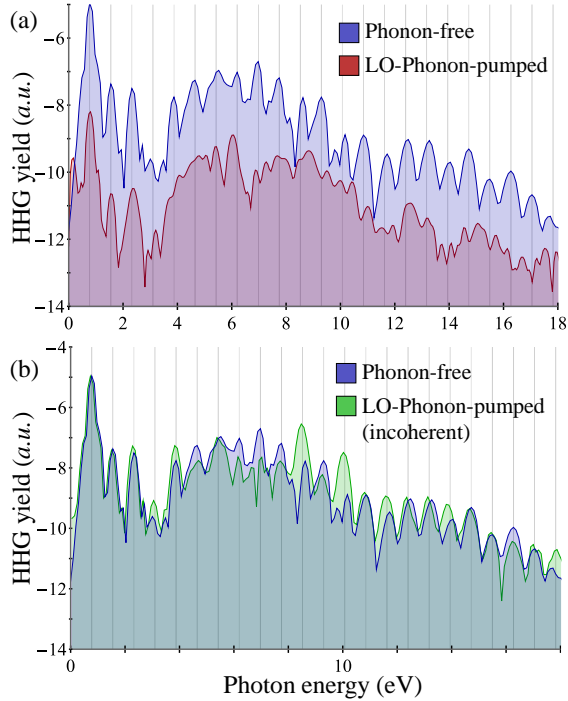

**Figure 2.** Effects of coherent phonons on HHG spectra. (a) HHG spectra with and without pre-excited LO phonon motion for IR laser polarized along the *x*-axis (transverse to phonon motion). The HHG spectra in both cases are shifted from one another for clarity. (b) HHG spectra from incoherent LO phonon case compared to the phonon-free response. The incoherent case was calculated by averaging over a full cycle of the pump-probe delay. Gray lines indicate integer harmonic orders of the IR photon energy, and all spectra are presented in log scale. For the calculation the LO mode was excited with an amplitude of 5.7% of the lattice parameter for bond stretching, and 4.95% of the lattice parameter for bond compression.

We next analyze the temporally-resolved response in the LO phonon case. Figure 3(a) and (b) present the pump-probe resolved HHG spectra for two probe laser polarizations (along the *x*-axis, and *y*-axis, respectively). As expected, the spectra are delay-dependent, and there are strong modulations of the HHG yield that oscillate periodically with the phonon period. Most notably, there is sharp polarization-sensitivity in this temporally-resolved response – if the IR probe laser is polarized along the active phonon mode, or transversely to it, the HHG yield is either enhanced or suppressed, respectively.

To further analyze these effects, we track a simple and more experimentally robust observable – the total integrated HHG yield in the plateau region. Figure 3(c) presents the integrated HHG yield *vs.* pump-probe delay, which oscillates periodically with a frequency $\Omega_{LO}$. The yield modulates by factors of 5-10 depending on the delay. Further analysis shows that the minimal yield is obtained when the peak of the probe laser envelope roughly coincides with the maximally-induced B-N bond stretching (that occurs at a pump-probe delay of ~10.1 fs). The minima is offset from the maximal bond-stretching by ~2.4 fs (it arises at a pump-probe delay of ~12.5 fs). Similarly, the maximal yield is obtained when the peak of the laser envelope roughly coincides with the maximally-induced B-N bond compression. When the probe laser polarization axis is rotated to be transverse to the phonon motion (along the *x*-axis), this dependence is exactly reversed (compare blue and red lines in Fig. 3(c)), which reflects the results in Figs. 3(a,b). This finding unambiguously establishes a real-space origin for the effect, because when the B-N bonds along the *y*-axis are maximally-compressed, those along the *x*-axis are maximally-stretched, and vise-versa. Thus, the real-space displacement of the ions is directly imprinted on the temporally-resolved HHG spectral response. A main result here is that the capability to temporally resolve the moment of maximal bond compression or stretching is limited by the duration and shape of the probe IR laser pulse, which in this case leads to the ~2.4 fs error. We also note slight deviations from the perfect matching of peaks and minimas in Fig. 3(c) at longer delays, which originate from higher order effects of electron-phonon and phonon-phonon couplings, as well as anharmonicity. These interesting phenomena are relatively small (<1 fs) such that they do not affect the main results derived here. We also highlight that there are other effects induced in the HHG spectra by the coherent phonons such as cutoff modulations, which might also prove useful for a similar spectroscopic analysis (see Fig. 3(a,b)).

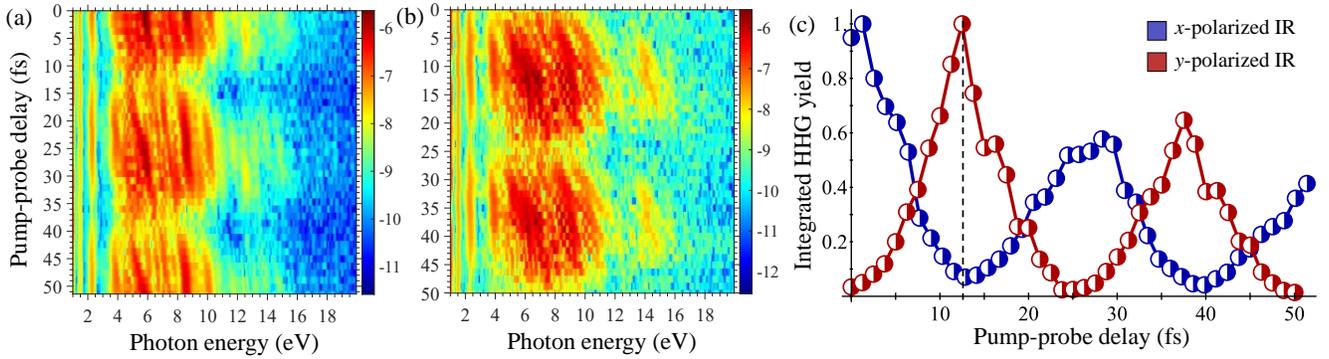

**Figure 3.** Temporally-resolved HHG response with active phonons. (a) HHG spectra vs. pump-probe delay for *x*-polarized IR probe pulse. (b) Same as (a), but for *y*-polarized IR probe pulse. The spectra are presented in log scale. (c) Normalized HHG yield integrated over the plateau region *vs.* the pump-probe delay for IR pulse polarized along the *x*-axis and *y*-axis. For the calculations the LO mode was excited with an amplitude of 5.7% of the lattice parameter for bond stretching, and 4.95% of the lattice parameter for bond compression. Dashed black line in (c) indicates the correspondence between the enhancement or suppression of the HHG yield, depending on the polarization axis of the IR pulse.

We further explore the origin of the strong oscillations in the HHG yield with pump-probe delay. Figure 4 presents the plateau-integrated HHG yield *vs.* the laser polarization angle with respect to the *x*-axis, which is calculated for three cases: (i) for the equilibrium configuration, (ii) for a static lattice where the ions are displaced to maximally compress the B-N bonds along the *y*-axis, (iii) for a static lattice where the ions are displaced to maximally stretch the B-N bonds along the *y*-axis. In general, the HHG yield from hBN is highly anisotropic, and stronger emission is obtained when the laser is polarized along the B-N bonds even in the equilibrium system, as well as along other high-symmetry axes (Fig. 4(a)). This result is in-line with previously observed anisotropies(46–49). When the lattice is distorted by few percent as in Fig. 4(b,c), this anisotropy is greatly enhanced, and the dominant HHG contributions arise when the laser polarization is parallel to the compressed B-N bond. This analysis verifies that the results in Fig. 3 indeed probe the instantaneous changes in the lattice structure (because the dominant effect in the temporal dynamics is reconstructed by just the static distorted lattice). It establishes an adiabatic picture for analyzing HHG yield modulations with respect to ultrafast structural distortions. We also note that an equivalent picture arises in *k*-space where the enhanced yield along the B-N bond can be rationalized due to instantaneous phonon-induced changes in the electronic band structure that couple to dominant interband HHG(2, 50) (see supplementary information (SI)). Similar results are obtained for the ZO mode, though here the dependence can be even stronger because ion displacement results in a large reduction of the band gap through formation of mid-gap states (see SI).

At this point it's worth mentioning that the strength of the effect (the depth of the HHG yield modulation with pump-probe delay) is connected to the amplitude of the excited phonon, where larger phonon amplitudes lead to more pronounced modulations (see SI). In particular, there is an exponential mapping between the HHG yield modulation and the pumped phonon amplitude. Moreover, the HHG cutoff is modulated with the pump-probe delay (see Fig. 3(a,b)). This clearly connects the emission to the instantaneous changes in the band structure and lattice geometry induced by the phonon motion, validating the adiabatic picture (see discussion in SI).

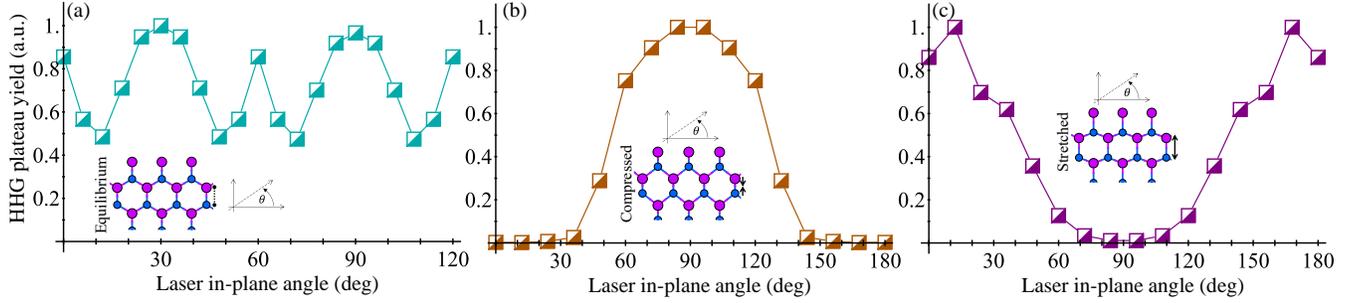

**Figure 4.** Polarization-resolved HHG response in hBN corresponding to the instantaneous lattice structure. (a) HHG plateau-integrated yield *vs.* the laser polarization axis for the static lattice in the equilibrium geometry (due to the 3-fold rotational symmetry only 120 degrees are presented). (b) Same as (a) but where the ions are statically frozen at a position with 4.95% bond compression for the B-N bond along the *y*-axis. (c) Same as (b) but where the ions are statically frozen at a position of 5.7% bond stretching for the B-N bond along the *y*-axis. (b) and (c) correspond to the maximally displaced lattice structures along the LO phonon dynamics in Fig. 3(c), where due to the reduced 2-fold rotational symmetry only 180-degrees are presented. Insets in all figures schematically represent the hBN structures (displacements of atoms in (b) and (c) are exaggerated for clarity), and the axis system denotes the angle of the laser in-plane polarization (from the *x*-axis in the geometries plotted in inset). Each plot is normalized to maximal power for clarity.

Up to this point, $\phi_{CEP}$ in eq. (1) was assumed to be fixed to zero. We now explore the option of performing CEP-stable experiments in a multi-dimensional spectroscopy configuration – recording the HHG response *vs.* the pump-probe delay, *vs.* the CEP. Figure 5(a) presents the calculated HHG response *vs.* CEP for the phonon-free system, while Fig. 5(b) presents the same spectra for one exemplary pump-probe delay in the LO phonon-pumped system. As seen in Fig. 5(a), there is virtually no CEP dependence in the HHG spectra in the phonon-free system. This corresponds to the common understanding that CEP-sensitivity should not arise from the multi-cycle pulses explored in our parameter regime. On the other hand, Fig. 5(b) shows that the HHG yield is modulated with the CEP when coherent phonons are present, especially in the plateau region. This establishes a new mechanism by which phonons can be probed through HHG-spectroscopy. We note that results of similar nature are obtained from other laser polarizations and from the ZO mode (see SI).

Interestingly, for our laser conditions a CEP shift of $\pi/2$ creates a temporal window just 1.3 fs wide (a quarter-cycle of the carrier wave period), which could offer enhanced selectivity towards the instantaneous lattice distortion (this temporal resolution is disconnected from the total pulse duration, and arises due to the CEP-sensitivity mechanism that is sub-cycle in nature). As an intuitive candidate for an observable that is capable of accessing this resolution (and can be readily measured in experiments), we consider the following observable – the normalized standard deviation of the plateau-integrated HHG yield with respect to the CEP (denoted as $\eta$). $\eta$ essentially quantifies the extent to which the HHG yield is CEP-sensitive for a given pump-probe delay. For instance, for the phonon-free system we find $\eta \approx 0.25\%$, i.e. there is virtually no CEP sensitivity, in accordance with Fig. 5(a). On the other hand, Fig. 5(c) presents $\eta$ *vs.* the pump-probe delay (blue line) for the LO-pumped system, which paints a different picture – there is relatively strong CEP dependence with $\eta \approx 4\%$ throughout, while for the particular delay of ~9.5 fs, there is a large peak with $\eta \approx 12\%$ (this strong effect should be observable even in systems with few percent experimental uncertainty).

Figure 5(c) validates that $\eta$ indeed provides enhanced sensing for the B-N bond compression – the peak at ~9.5 fs occurs at the pump-probe delay for which the phonon-induced maximal bond-compression almost coincides with the peak of the pulse envelope (the temporal distance between the peak of the laser envelope and the phonon-induced maximal bond compression is denoted by $\delta t$ and shown in green in Fig. 5(c)). In fact, the maximal bond compression is obtained at 10.1 fs, leading to an error of just 0.6 fs (this can be compared to the peak in the HHG yield *vs.* delay in Fig. 3(c), which was offset by 2.4 fs, a factor 4 larger). The CEP-resolved spectra thus provides additional temporal information that was not present in the delay-dependent spectra, which exhibits enhanced selectivity and temporal resolution. Consequently, the multi-dimensional spectra can be used to 'set the clock' to the exact moment of bond compression. This type of pump-probe-CEP multi-dimensional spectroscopy therefore offers enhanced sensing for ultrafast lattice distortions with resolutions of ~1 fs. We also note in Fig. 5(c) the appearance of a minor secondary peak at a pump-probe delay of ~12.2 fs, that corresponds more closely to the one in Fig. 3(c). The origin of this peak remains unclear, but we hypothesize that it corresponds either to slightly delayed electron dynamics driven in the system, or to a more subtle change in the electronic structure of the material during the bond stretching process. The exact nature of the effect should be explored in future work. Importantly, it highlights the ability of the new multi-dimensional observable, $\eta$, to capture novel phenomena. Lastly, we highlight the extremely nonlinear nature of the new multi-dimensional CEP-based spectroscopy technique – $\eta$ in Fig. 5(c) does not show a peak at bond stretching that is equivalent to the one obtained upon bond compression. In that respect, the curve is highly nonlinear with particular sensitivity for the compression process. Thus, the new approach might be able to tap into the extreme nonlinearly of the HHG process for probing other processes as well.

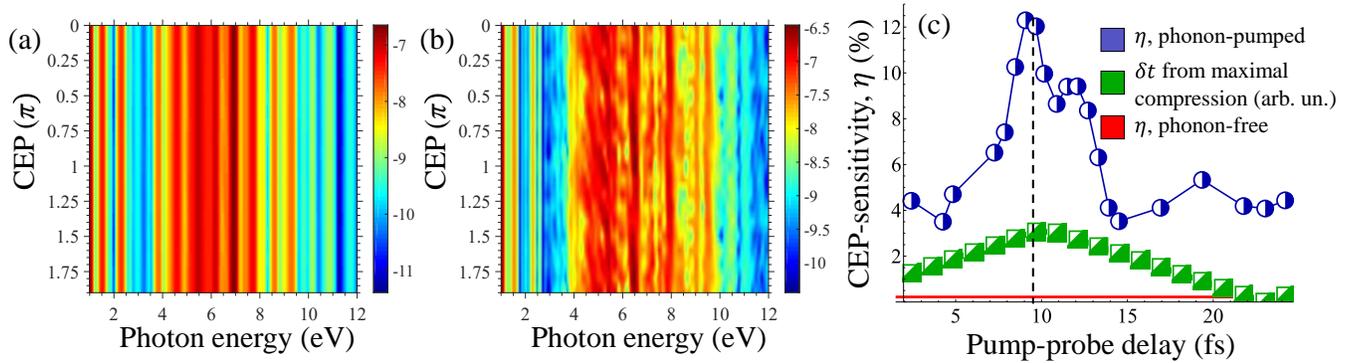

**Figure 5.** CEP-dependent HHG response for LO driven coherent phonon dynamics. (a) CEP-dependent HHG spectra for the phonon-free static lattice in the equilibrium geometry (in log scale). No CEP-dependence is present in the phonon-free case. (b) Same as (a) but for the LO-pumped system for the particular pump-probe delay of 10.1 fs, showing onset of CEP-sensitivity. (c) Degree of CEP-sensitivity ($\eta$) in the phonon-pumped system *vs.* pump-probe delay (blue line), on the same scale as the temporal distance between the phonon-induced instantaneous maximal B-N bond compression, and the peak of the laser envelope, $\delta t$ (green, given in the plot in arb. un.). The IR probe laser polarization is along the *x*-axis, and the LO phonon is excited just an in Fig. 3. The red line indicates $\eta$ for the phonon-free system. The dashed black line indicates the correspondence between the moment of maximal bond compression, and the peak in $\eta$.

**Conclusions**

To conclude, through *ab-initio* calculations in hBN, we explored the possibility to probe phonons with pump-probe HHG measurements. We predicted several experimentally detectable effects that arise from coherently-pumped phonons: (i) the HHG spectrum can form a quasi-continuous emission rather than discrete harmonic peaks due to the phonon timescale. We directly showed that this effect vanishes for incoherent phonons. (ii) The harmonic yield in the plateau region is strongly delay-dependent, and oscillates periodically in accordance with the instantaneous structural changes in the lattice such as bond compression and stretching. (iii) If the phonon motion has a timescale on the same order of magnitude as the pulse duration, CEP-sensitivity emerges in the HHG spectra, which can be measured *vs.* the pump-probe delay for an enhanced sensing of lattice distortions. Most importantly, the temporal resolution achievable *via* the CEP-multi-dimensional spectroscopy is on the order of ~1 femtosecond, and is well below the total pulse duration, because it is inherently connected with sub-cycle dynamics (unlike standard pump-probe spectroscopy that is limited by the probe pulse duration and shape).

While we limited our analysis to monolayer hBN (because it has relatively fast phonons that are easier to simulate and analyze), the discussed physical mechanisms are general and should apply to other materials as well, including bulk materials. In particular, it is possible that some of the CEP-sensitivity observed in ref.(14) originated from a similar underlying mechanism (because in that work a THz pump was used that could excite surface phonon motion). Moreover, we emphasize that the phonon pumping method is versatile, and one could imagine also implementing this technique by pumping phonons with IR pulses as well as other excitation schemes (including phonons with finite momenta). All of these phenomena present exciting prospects for controlling the HHG emission, as well as for the application of multi-dimensional HHG-spectroscopy for probing phonons, lattice structural distortions, and phase transitions. Looking forward, we expect that our approach might also be applied for probing nonlinear phononics(22, 43), phonon-polariton motion in real-time(51, 52), hyperbolic materials(53, 54), and real-time probing of electron-phonon coupling with possible applications for superconductivity.

**Computational details**

We report here on technical details for the calculations presented in the main text. We start by presenting our methodological approach that is based on time-dependent density functional theory (TDDFT). All DFT calculations were performed using the real-space grid-based code, octopus(36–38). The Kohn Sham (KS) equations were discretized on a Cartesian grid with the shape of the primitive lattice cell, where equilibrium atomic geometries and lattice parameters were taken at the experimental values. The $z$-axis (transverse to the hBN monolayer) was described using non-periodic boundaries with a length of 50 Bohr. Calculations were performed using the local density approximation (LDA) for the exchange-correlation (XC) functional. Spin degrees of freedom and spin-orbit couplings were neglected. The frozen core approximation was used for core levels which were treated with appropriate norm-conserving pseudopotentials(55). The KS equations were solved to self-consistency with a tolerance $<10^{-7}$ Hartree, and the grid spacing was converged to $\Delta x=\Delta y=\Delta z=0.3$ Bohr. We employed a $\Gamma$-centered 36×36×1 $k$-grid, which converged the HHG spectrum and the phonon trajectories. We note that the relatively small grid spacing of 0.3 Bohr was required to obtain converged phonon trajectories.

For TDDFT calculations, we solved the time-dependent KS equations within the adiabatic approximation, represented in real-space and in the velocity gauge, given in atomic units by:

$$i\partial_t |\varphi_{n,k}^{KS}(t)\rangle = \left(\frac{1}{2}\left(-i\nabla + \frac{\mathbf{A}(t)}{c}\right)^2 + v_{KS}(\mathbf{r},t)\right)|\varphi_{n,k}^{KS}(t)\rangle \quad (2)$$

where $|\varphi_{n,k}^{KS}(t)\rangle$ is the KS-Bloch state at $k$-point $k$ and band $n$, $\mathbf{A}(t)$ is the vector potential of the laser electric field within the dipole approximation, such that $-\partial_t \mathbf{A}(t) = c\mathbf{E}(t)$, $c$ is the speed of light in atomic units ($c\approx137.036$), and $v_{KS}(\mathbf{r},t)$ is the time-dependent KS potential given by:

$$v_{KS}(\mathbf{r},t) = -\sum_I \frac{Z_I}{|\mathbf{R}_I(t) - \mathbf{r}|} + \int d^3r' \frac{n(\mathbf{r}',t)}{|\mathbf{r}-\mathbf{r}'|} + v_{XC}[n(\mathbf{r},t)] \quad (3)$$

where $Z_I$ is the charge of the $I$'th nuclei and $\mathbf{R}_I(t)$ is its coordinate, $v_{XC}$ is the XC potential that is a functional of $n(\mathbf{r},t)=\sum_{n,k}\left||\varphi_{n,k}^{KS}(t)\rangle\right|^2$, the time-dependent electron density. The KS wave functions were propagated with a time step of $\Delta t$=0.2 a.u. which converged the HHG spectra. The initial state was taken to be the system's ground state at the equilibrium geometry (i.e. $\mathbf{R}_I(t=0)$ describes the lattice at equilibrium), which assumes that the Born-Oppenheimer approximation is valid prior to the turn-on of the IR laser pulse. The propagator was represented by a Lanczos expansion. In the time-dependent calculations, we employed absorbing boundaries through complex absorbing potentials (CAPs) along the periodic $z$-axis with a width of 12 Bohr, and ionization was well below 1% in all calculations.

The KS TDDFT equations of motion were coupled to classical phonons by the time-dependent ionic positions $\mathbf{R}_I(t)$. The ion positions were allowed to evolve dynamically, and **were** calculated by Ehrenfest dynamics as implemented in octopus code, where forces acting on the ions are evaluated directly from the electronic density and the laser(39). The initial condition for $\mathbf{R}_I(t)$ was taken as the equilibrium geometry of the lattice, but where the ions were given initial velocities according to the particular phonon eigenmode (we verified that this approach is equivalent to starting the motion by displacing the ions directly and initiating the dynamics without an initial velocity). For the phonon-free calculations $\mathbf{R}_I$ was kept frozen as is standardly done in TDDFT, either at the equilibrium geometry, or at the distorted geometry as presented in the main text. We note that small errors in the ion trajectories were numerically observed, causing a drift of the hBN monolayer in space by up to 0.1 Bohr over 50 fs. We have verified that these errors arise due to the grid representation, which leads to small errors in the calculated ionic forces that build-up over time for the long propagation durations. The contribution of this type of error to the HHG spectra was verified to be negligible, since the total drift of the lattice in our conditions is small and slow.

The time-dependent current expectation value was calculated directly from the time-dependent KS states as:

$$\mathbf{J}(t) = \frac{1}{\Omega} \int_\Omega d^3 r\, \mathbf{j}(\mathbf{r}, t) \tag{4}$$

where $\mathbf{j}(\mathbf{r}, t)$ is the microscopic time-dependent current density:

$$\mathbf{j}(\mathbf{r}, t) = \frac{1}{2} \sum_{n,k} \left[ \varphi_{n,k}^{KS\,*}(r,t) \left( -i\nabla + \frac{\mathbf{A}(t)}{c} \right) \varphi_{n,k}^{KS}(r,t) + c.c. \right] \tag{5}$$

, and $\Omega$ represents the volume integral over the primitive cell. Note that the ionic current is neglected throughout this work, because it is expected to only dominantly contribute at the phonon resonance frequency. The HHG spectrum was calculated as the Fourier transform of the first derivative of the current:

$$I(\omega) = \left| \int dt\, \partial_t \{ f(t)\, \mathbf{J}(t) \} e^{-i\omega t} \right|^2 \tag{6}$$

, which was evaluated numerically with an 8'th order finite-difference approximation for the temporal derivative, and fast Fourier transforms. The HHG spectra in the paper only present the harmonic components within the hBN *xy*-plane that propagate to the far field (i.e. any residual *z*-currents that occur when there is ZO phonon motion are not plotted). Note that prior to calculating the HHG yield, the current was filtered with a temporal window similar to the laser field envelope (as in eq. (6) above). This procedure greatly reduces any noise in the emission due to the phonon motion and suppresses the phonon-resonant peak.

The envelope function of the employed laser pulse, *f(t)* from eq. (1), was taken to be of the following 'super-sine' form(56):

$$f(t) = \left( \sin\left( \pi \frac{t}{T_p} \right) \right)^{\left( \frac{\left| \pi \left( \frac{t}{T_p} - \frac{1}{2} \right) \right|}{\sigma} \right)} \tag{7}$$

where $\sigma=0.75$, $T_p$ is the duration of the laser pulse which was taken to be $T_p=8T$ (~25 fs full-width-half-max (FWHM)), where *T* is a single cycle of the fundamental carrier frequency that corresponds to 1600nm light. This form is roughly equivalent to a super-gaussian pulse, but where the field starts and ends exactly at zero amplitude, which is more convenient numerically.

## Acknowledgments

We acknowledge financial support from the European Research Council (ERC-2015-AdG-694097). The Flatiron Institute is a division of the Simons Foundation. O.N. gratefully acknowledges the generous support of the Humboldt foundation, and a Schmidt Science Fellowship. J.Z. acknowledges funding from the European Union's Horizon 2020 research and innovation program under the Marie Sklodowska-Curie grant agreement No. 886291 (PeSD-NeSL). This work was supported by the Cluster of Excellence Advanced Imaging of Matter (AIM), Grupos Consolidados (IT1249-19) and SFB925.

## References

1. S. Ghimire, *et al.*, Observation of high-order harmonic generation in a bulk crystal. *Nat. Phys.* **7**, 138–141 (2011).
2. S. Ghimire, D. A. Reis, High-harmonic generation from solids. *Nat. Phys.* **15**, 10–16 (2019).
3. S. Sederberg, *et al.*, Vectorized optoelectronic control and metrology in a semiconductor. *Nat. Photonics* **14**, 680–685 (2020).
4. G. Vampa, *et al.*, All-Optical Reconstruction of Crystal Band Structure. *Phys. Rev. Lett.* **115**, 193603 (2015).
5. A. A. Lanin, E. A. Stepanov, A. B. Fedotov, A. M. Zheltikov, Mapping the electron band structure by intraband high-harmonic generation in solids. *Optica* **4**, 516–519 (2017).
6. H. Liu, *et al.*, High-harmonic generation from an atomically thin semiconductor. *Nat. Phys.* **13**, 262–265 (2017).

7. N. Saito, *et al.*, Observation of selection rules for circularly polarized fields in high-harmonic generation from a crystalline solid. *Optica* **4**, 1333–1336 (2017).
8. O. Neufeld, D. Podolsky, O. Cohen, Floquet group theory and its application to selection rules in harmonic generation. *Nat. Commun.* **10**, 405 (2019).
9. T. T. Luu, H. J. Wörner, Measurement of the Berry curvature of solids using high-harmonic spectroscopy. *Nat. Commun.* **9**, 916 (2018).
10. D. Bauer, K. K. Hansen, High-harmonic generation in solids with and without topological edge states. *Phys. Rev. Lett.* **120**, 177401 (2018).
11. R. E. F. Silva, Á. Jiménez-Galán, B. Amorim, O. Smirnova, M. Ivanov, Topological strong-field physics on sub-laser-cycle timescale. *Nat. Photonics* **13**, 849–854 (2019).
12. Y. Bai, *et al.*, High-harmonic generation from topological surface states. *Nat. Phys.* **17**, 311–315 (2021).
13. D. Baykusheva, *et al.*, Strong-field physics in three-dimensional topological insulators. *Phys. Rev. A* **103**, 23101 (2021).
14. C. P. Schmid, *et al.*, Tunable non-integer high-harmonic generation in a topological insulator. *Nature* **593**, 385–390 (2021).
15. D. Baykusheva, *et al.*, All-Optical Probe of Three-Dimensional Topological Insulators Based on High-Harmonic Generation by Circularly Polarized Laser Fields. *Nano Lett.* **21**, 8970–8978 (2021).
16. O. Schubert, *et al.*, Sub-cycle control of terahertz high-harmonic generation by dynamical Bloch oscillations. *Nat. Photonics* **8**, 119–123 (2014).
17. T. T. Luu, *et al.*, Extreme ultraviolet high-harmonic spectroscopy of solids. *Nature* **521**, 498–502 (2015).
18. N. Yoshikawa, T. Tamaya, K. Tanaka, Optics: High-harmonic generation in graphene enhanced by elliptically polarized light excitation. *Science* **356**, 736–738 (2017).
19. Z. Wang, *et al.*, The roles of photo-carrier doping and driving wavelength in high harmonic generation from a semiconductor. *Nat. Commun.* **8**, 1686 (2017).
20. A. J. Uzan, *et al.*, Attosecond spectral singularities in solid-state high-harmonic generation. *Nat. Photonics* (2020) https:/doi.org/10.1038/s41566-019-0574-4.
21. M. R. Bionta, *et al.*, Tracking ultrafast solid-state dynamics using high harmonic spectroscopy. *Phys. Rev. Res.* **3**, 023250 (2021).
22. J. S. Ginsberg, *et al.*, Optically-Induced Symmetry Breaking via Nonlinear Phononics. *arXiv:2107.11959* (2021).
23. S. Wall, *et al.*, Ultrafast changes in lattice symmetry probed by coherent phonons. *Nat. Commun.* **3**, 721 (2012).
24. H. Hübener, U. De Giovannini, A. Rubio, Phonon Driven Floquet Matter. *Nano Lett.* **18**, 1535–1542 (2018).
25. D. Shin, *et al.*, Phonon-driven spin-Floquet magneto-valleytronics in MoS2. *Nat. Commun.* **9**, 638 (2018).
26. F. Grandi, J. Li, M. Eckstein, Ultrafast Mott transition driven by nonlinear electron-phonon interaction. *Phys. Rev. B* **103**, L041110 (2021).
27. M.-F. Lin, *et al.*, Ultrafast non-radiative dynamics of atomically thin MoSe2. *Nat. Commun.* **8**, 1745 (2017).
28. N. L. Wagner, *et al.*, Monitoring molecular dynamics using coherent electrons from high harmonic generation. *Proc. Natl. Acad. Sci.* **103**, 13279 LP – 13285 (2006).
29. A. Ferré, *et al.*, Multi-channel electronic and vibrational dynamics in polyatomic resonant high-order harmonic generation. *Nat. Commun.* **6**, 5952 (2015).
30. L. He, *et al.*, Monitoring ultrafast vibrational dynamics of isotopic molecules with frequency modulation of high-order harmonics. *Nat. Commun.* **9**, 1108 (2018).
31. S. T. Cundiff, Phase stabilization of ultrashort optical pulses. *J. Phys. D. Appl. Phys.* **35**, R43–R59 (2002).
32. A. Baltuška, *et al.*, Attosecond control of electronic processes by intense light fields. *Nature* **421**, 611–615 (2003).
33. M. Nisoli, *et al.*, Effects of Carrier-Envelope Phase Differences of Few-Optical-Cycle Light Pulses in Single-Shot High-Order-Harmonic Spectra. *Phys. Rev. Lett.* **91**, 213905 (2003).
34. O. Neufeld, A. Fleischer, O. Cohen, High-order harmonic generation of pulses with multiple timescales: selection rules, carrier envelope phase and cutoff energy. *Mol. Phys.* **117**, 1956–1963 (2019).
35. O. Neufeld, N. Tancogne-Dejean, U. De Giovannini, H. Hübener, A. Rubio, Light-Driven Extremely Nonlinear Bulk Photogalvanic Currents. *Phys. Rev. Lett.* **127**, 126601 (2021).
36. A. Castro, *et al.*, octopus: a tool for the application of time-dependent density functional theory. *Phys. status solidi* **243**, 2465–2488 (2006).
37. X. Andrade, *et al.*, Real-space grids and the Octopus code as tools for the development of new simulation approaches for electronic systems. *Phys. Chem. Chem. Phys.* **17**, 31371–31396 (2015).
38. N. Tancogne-Dejean, *et al.*, Octopus, a computational framework for exploring light-driven phenomena and quantum dynamics in extended and finite systems. *J. Chem. Phys.* **152**, 124119 (2020).
39. X. Andrade, *et al.*, Modified Ehrenfest Formalism for Efficient Large-Scale ab initio Molecular Dynamics. *J. Chem. Theory Comput.* **5**, 728–742 (2009).
40. M. A. L. Marques, *et al.*, "Time-Dependent Density Functional Theory" in *Time-Dependent Density Functional Theory*, (Springer, 2003).
41. C. Hartwigsen, S. Goedecker, J. Hutter, Relativistic separable dual-space Gaussian pseudopotentials from H to Rn. *Phys. Rev. B* **58**, 3641–3662 (1998).
42. M. Först, *et al.*, Nonlinear phononics as an ultrafast route to lattice control. *Nat. Phys.* **7**, 854–856 (2011).
43. F. Iyikanat, A. Konečná, F. J. García de Abajo, Nonlinear Tunable Vibrational Response in Hexagonal Boron Nitride. *ACS Nano* (2021) https://doi.org/10.1021/acsnano.1c03775.

44. Á. Jiménez-Galán, R. E. F. Silva, O. Smirnova, M. Ivanov, Lightwave control of topological properties in 2D materials for sub-cycle and non-resonant valley manipulation. *Nat. Photonics* **14**, 728–732 (2020).
45. Á. Jiménez-Galán, R. E. F. Silva, O. Smirnova, M. Ivanov, Sub-cycle valleytronics: control of valley polarization using few-cycle linearly polarized pulses. *Optica* **8**, 277–280 (2021).
46. H. Liu, *et al.*, Enhanced high-harmonic generation from an all-dielectric metasurface. *Nat. Phys.* **14**, 1006–1010 (2018).
47. Y. S. You, E. Cunningham, D. A. Reis, S. Ghimire, Probing periodic potential of crystals via strong-field re-scattering. *J. Phys. B At. Mol. Opt. Phys.* **51**, 114002 (2018).
48. Y. S. You, D. A. Reis, S. Ghimire, Anisotropic high-harmonic generation in bulk crystals. *Nat. Phys.* **13**, 345–349 (2017).
49. H. Lakhotia, *et al.*, Laser picoscopy of valence electrons in solids. *Nature* **583**, 55–59 (2020).
50. M. Wu, D. A. Browne, K. J. Schafer, M. B. Gaarde, Multilevel perspective on high-order harmonic generation in solids. *Phys. Rev. A* **94**, 1–13 (2016).
51. N. Li, *et al.*, Direct observation of highly confined phonon polaritons in suspended monolayer hexagonal boron nitride. *Nat. Mater.* **20**, 43–48 (2021).
52. Y. Kurman, *et al.*, Spatiotemporal imaging of 2D polariton wave packet dynamics using free electrons. *Science* **372**, 1181 LP – 1186 (2021).
53. G. Hu, J. Shen, C.-W. Qiu, A. Alù, S. Dai, Phonon Polaritons and Hyperbolic Response in van der Waals Materials. *Adv. Opt. Mater.* **8**, 1901393 (2020).
54. S. Dai, *et al.*, Tunable Phonon Polaritons in Atomically Thin van der Waals Crystals of Boron Nitride. *Science* **343**, 1125 LP – 1129 (2014).
55. C. Hartwigsen, S. Goedecker, J. Hutter, Relativistic separable dual-space Gaussian pseudopotentials from H to Rn. *Phys. Rev. B* **58**, 3641–3662 (1998).
56. O. Neufeld, O. Cohen, Background-Free Measurement of Ring Currents by Symmetry-Breaking High-Harmonic Spectroscopy. *Phys. Rev. Lett.* **123**, 103202 (2019).

# Supplementary Information

**Additional results for ZO phonon mode.** We present here additional results complementary to those presented in the main text but for the ZO phonon-pumped system. Figure S1(a) presents the delay dependent HHG yield for ZO phonon amplitude of 11% out-of-plane motion with respect to the lattice parameter (note that the variation in the B-N bond lengths is much smaller due to the planar geometry of the lattice and amounts to ~1.8% in this case). Similar periodic dependence is observed, corroborating the results presented in the main text for the LO mode. Here the peaks and minima correspond to the lattice phonon-induced structural changes, where the ions are maximally stretched out-of-plane near the peak HHG yield, and are near the equilibrium position at the minimum emission. We also note that the HHG yield is modulated twice per phonon period due to the symmetry of this mode's motion – the HHG response is identical whether the N atoms move above the layer and the B atoms below it, or vice versa.

Figure S1(b) presents exemplary CEP-dependent spectra for the ZO driven case (in this case with a phonon amplitude of 19.3% of the lattice parameter for out-of-plane motion, corresponding to 5.4% maximal B-N bond stretching), showing that a similar CEP-sensitivity arises as was demonstrated in the main text. The emerging CEP sensitivity is even more pronounced for the ZO mode due to the larger changes it induces in the instantaneous band structure (as we show below). We have verified that similar in nature results are obtained also for other laser polarization and phonon amplitudes (not presented), validating the generality of the discussion in the main text.

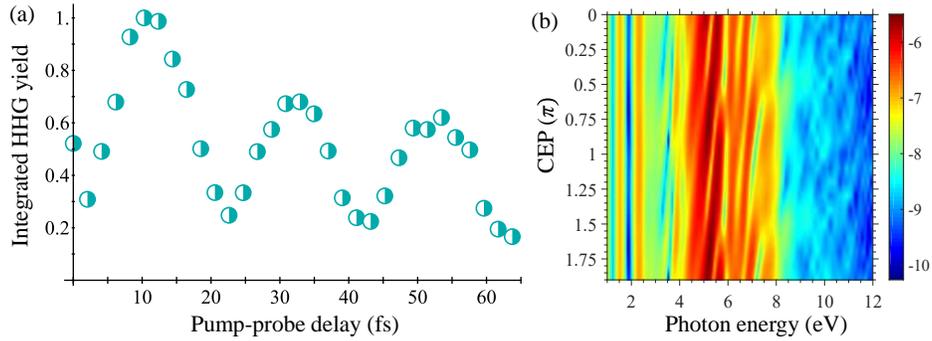

**Fig. S1.** HHG from ZO Phonon-pumped system. (a) Delay-dependent HHG plateau yield for out-of-pane phonon amplitude of 11% of the lattice parameter (1.8% bond stretching), and laser polarization along the *x*-axis. (b) Exemplary CEP-dependent HHG spectra for the ZO-pump system, in similar conditions to (a), but with a pump-probe delay of 9.3 femtoseconds and phonon amplitude of 19.3% for the out-of-plane motion (5.4% bond stretching). The spectra is presented in log scale.

**Phonon-induced band structure changes.** We present here an analysis of the phonon-induced changes to the electronic band-structure within a Born-Oppenheimer adiabatic picture. This complements the real-space analysis presented in the main text for the origin of the strong HHG selectivity to variations in the instantaneous B-N bond length. Figure S2(a) presents the LDA-calculated band structure of hBN (KS eigenvalues) along selected high symmetry lines. Note that the plot includes the formally equivalent high symmetry points k={0.5,0} (the *M* point) and k={0,0.5} (denoted as *M''*), because these points become non-equivalent in the distorted lattice that is no longer 3-fold symmetric (with a reduced instantaneous 2-fold symmetry for LO induced changes). In equilibrium, both the valence band maxima (VBM) and conduction band minima (CBM) are positioned at the *K* point, with a direct gap of 4.51 eV. Figures S2(b,c) present the calculated band structures on a similar level of theory, but where the ions have been displaced to their extrema position along the LO phonon trajectory (i.e. at maximal B-N bond compression and stretching, respectively, at the phonon amplitude of 5.7% (corresponding to the figures in the main text)). Several observations can be noted. Firstly, the band gap and positions of the VBM and CBM are different for bond stretching and compression – the CBM shifts to the *M* point for bond compression, or to the *M''* point for bond stretching. This is accompanied by a significant direct band gap closing at the *Γ* point, which is energetically very close to the VBM in both cases. Secondly, in both cases the direct optical gap is diminished compared to the equilibrium lattice, but it is more strongly reduced in the compressed structure (up to 4.34 eV).

The disparity between the band gap changes alone are not enough to explain the enhanced selectivity towards bond compression. This result could potentially be attributed to the induced changes in the directionality of the band structure. To further analyze the mechanism in $k$-space, we recall that the high harmonics in the energy region of ~10eV are above the band gap, and are likely generated by interband transitions with recombination surrounding the $\Gamma$ point region (because those are the only regions allowing for high energy emission below the maximal band gap, see Fig. S2). This point of view assumes for simplicity that the main contribution to the plateau emission arises from the first valence and conduction bands. In the equilibrium lattice, this suggests that there should be stronger emission along lasers polarized from $K$ to $\Gamma$ (between the bonds, along the $x$-axis), because the initial excitation is preferred at the minimal band gap at the $K$-point, but for efficient HHG the conduction electrons need to be accelerated to regions near $\Gamma$ (because those are the only regions with direct gaps of ~10eV in the plateau region). Similarly, emission for laser polarization from $M$ to $\Gamma$ (along the bonds, along the $y$-axis) should also be intense, because the direct gap at the $M$ point is just a little bit higher than at $K$. This analysis fits well with the results in Fig. 4(a) in the main text, and thus we apply it also to the distorted lattices. For bond compression, the minimal optical gaps shifts to the M point, suggesting stronger emission along the $y$-axis. For bond stretching the minimal optical gap remains at the K, but is almost identical at the M'', both of which suggest stronger emission for lasers polarized along the $x$-axis. Of course, this simple picture is not quantitative and neglects other important features such as density of states and other regions in k-space that are not along high symmetry lines. Nevertheless, it roughly describes the $k$-space picture for the enhanced selectivity for just one of these processes depending on the laser polarization axis.

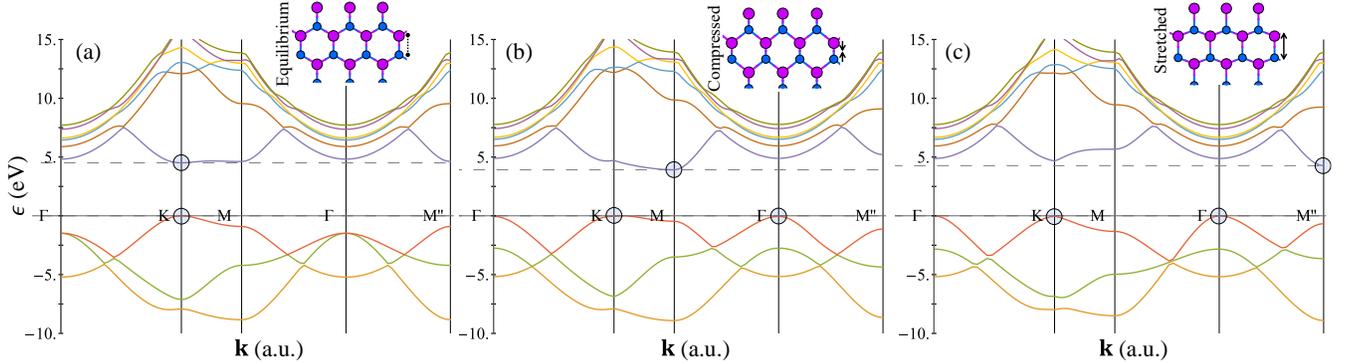

**Fig. S2.** LO Phonon-induced band-structure changes. (a) Equilibrium band-structure along high symmetry lines, calculated within LDA. (b) Same as (a), but where the ions are displaced by 5.7% towards compression of the y-axis B-N bonds. (c) Same as (b) but for bond stretching of 4.95%. The highest occupied and lowest unoccupied levels are denoted with horizontal dashed lines, the VBM and CBM are denoted with blue circles, and insets represent the lattice geometry.

Figure S3 presents a similar analysis for the ZO mode. The main difference here compared to the LO mode is that this mode always stretches the BN bonds, and thus a contrast is imprinted onto the HHG response between the maximal stretching, and the equilibrium geometry. Notably, the ZO mode breaks other lattice symmetries than the LO mode, and allows for a mid-gap state to be generated that greatly reduces the band gap. As a result of this, there can be much larger HHG yield variations for the ZO-mode, and even enhanced CEP-sensitivity that is especially sensitive towards the formation of this mid-gap state at moment of maximal stretching. The HHG mechanism in the presence of the mid-gap has not been explored in this work, and will be topic of future studies.

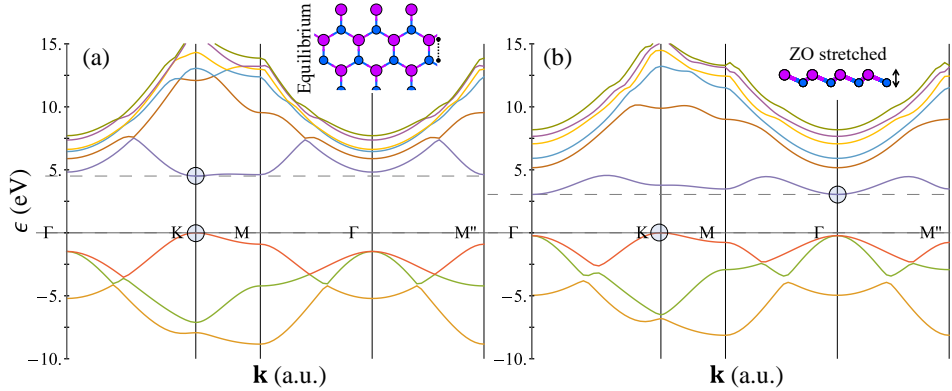

**Fig. S3.** ZO Phonon-induced band-structure changes. (a) Equilibrium band-structure along high symmetry lines, calculated within LDA. (b) Same as (a), but where the ions are displaced by 5.4% towards the maximal out-of-plane bond stretching. The highest occupied and lowest unoccupied levels are denoted with horizontal dashed lines, the VBM and CBM are denoted with blue circles, and insets represent the lattice geometry.

**HHG yield *vs.* phonon amplitude.** We explore here the HHG yield modulation (the contrast obtained in Fig. 3(c) in the main text *vs.* pump probe delay) with respect to the pumped-phonon amplitude. Fig. S4 presents data of the integrated HHG yield in the first plateau region both from LO and ZO phonon modes. A clear exponential scaling is obtained in both cases (with $R^2$ values of 0.9954 and 0.9992, for the LO and ZO modes, respectively). This indicates that the mechanism behind the HHG yield modulation is connected to changes in the instantaneous electronic structure that is adiabatically attenuated with the lattice geometry (because changes in the band structure would map exponentially to the HHG yield due to the inherent extreme nonlinearity). The exact mechanism behind this effect is not yet known, but will be topic of future studies. We hypothesize that it is connected to local changes in the electron density of sigma bonded electrons (in the B-N bonds) which might become more mobile under bond compression, or less under bond stretching.

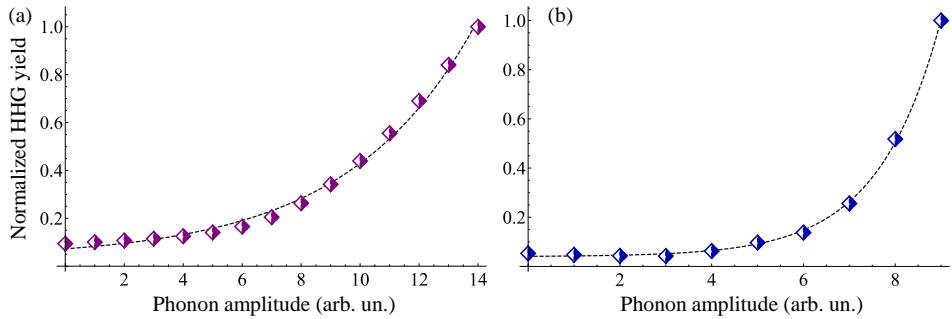

**Fig. S4.** HHG yield modulation *vs.* phonon amplitude. (a) Calculations for LO mode in hBN in the same conditions as Fig. 2(a) in the main text, but where the phonon amplitude is increased. (b) Same as (a) but for the ZO mode with conditions similar to those in Fig. S1. Dashed lines indicate exponential function best fits.

**Polarization sensitivity of perturbative harmonics.** For completeness, we include here results similar to those in Fig. 4 in the main text, but for the polarization modulated harmonic power in the perturbative harmonics (harmonics 2, 3, and 4). Fig. S5 presents the integrated harmonic yield in the perturbative region (excluding the linear response peak). A similar trend to that in Fig. 4 in the main text is seen also in the perturbative regime. This is regardless of the fact that in monolayer hBN the perturbative harmonics are only weakly modulated with the laser polarization axis in the equilibrium geometry (at least in this examined regime). This indicates that the mechanism that modulates the HHG yield due to the phonon motion is general. Notably, the contrast in the yield modulation *vs.* laser orientation is slightly smaller for the perturbative harmonics compared to the higher harmonics, indicating that the extreme nonlinearity is playing an additional role.

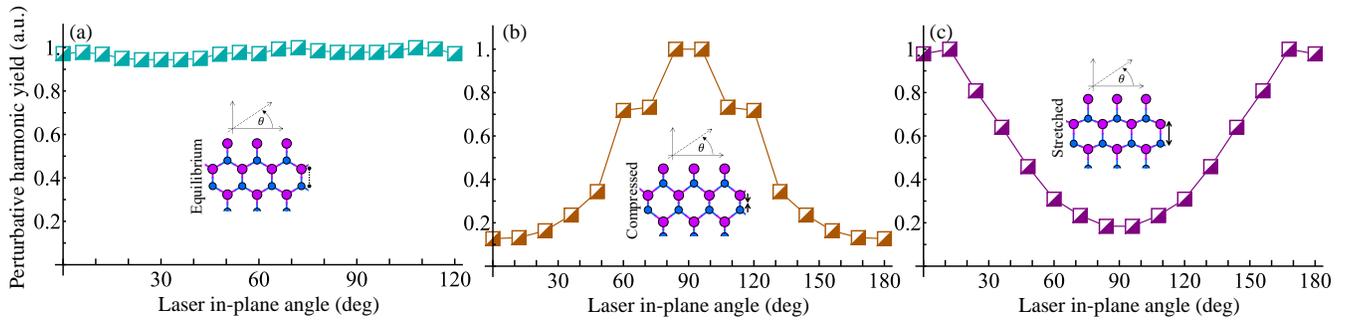

**Fig. S5.** Perturbative harmonic yield modulation *vs.* laser in-plane orientation in the static case for: (a) the equilibrium lattice, (b) the compressed lattice, and (c) the stretched lattice. The lattice geometries and laser conditions correspond to those in Fig. 4 in the main text. The insets correspond to the insets in Fig. 4 in the main text.

15